*Putnam's Critical and Explanatory Tendencies Interpreted from a Machine Learning Perspective*

Sheldon Z. Soudin

Introduction

Making sense of theory choice in normal and across extraordinary science is central to philosophy of science. The emergence of machine learning models has the potential to act as a wrench in the gears of current debates. In this paper, I will attempt to reconstruct the main movements that lead to and came out of Putnam's critical and explanatory tendency distinction, argue for the biconditional necessity of the tendencies, and conceptualize that wrench through a machine learning interpretation of my claim. Some preliminary definitions and statements of assumptions are in order.

Kuhn's picture of normal versus extraordinary science is presented in his 1962 book "The Structure of Scientific Revolution". In a short caricature of the distinction, normal science takes place within paradigms and extraordinary science takes place across paradigms. As such, extraordinary science entails scientific revolutions and paradigm shifts. The term "paradigm" thus becomes an important term for Kuhn's argument; however, it remains relatively ambiguously defined. For the purposes of this paper a paradigm may be reduced to established scientific theories, symbolic generalizations, and heuristic models. As a response to Kuhn's and Popper's positions on the nature of good theory choice, Putnam constructs schemata to illustrate two tendencies in the consideration of scientific problems. (The 'Corroboration' of Theories)

The critical tendency:

" <u>SCHEMA I</u>

<u>THEORY</u>

<u>AUXILIARY STATEMENTS</u>

PREDICTION – TRUE OR FALSE?     " (Putnam, 1979)

The explanatory tendency:

" <u>SCHEMA II</u>

THEORY

<u>??????????</u>

FACT TO BE EXPLAINED    " (Putnam, 1979)

Putnam endeavors to argue that Popper's falsifiability criterion is captured by schema I, and that theories alone cannot predict anything. It is rather the conjunction of theory and auxiliary statements that make a prediction. He suggests the explanatory tendency better captures the process of theory choice that occurs in normal science.

Machine learning, and more specifically deep learning is a computer science and applied mathematics method that constructs neural networks, trains it on a given data set and ultimately outputs recreations/predictions that emulate the data set, upon which it was trained. The specific training methods and modes of application vary, from convolutional neural networks to recurrent neural networks, from image classification to large language models (i.e. ChatGPT). For the sake of this paper and argumentation, I assume that neural networks are scalable, that is that increasing the number of computational resources in the model's training directly result in higher accuracy in the model's predictions. Two additional concepts are pertinent; abstractly put, parameters are the numerical values of neural networks. As such, it is these that are learned when a computer goes through the energetically consuming process of training a large language model for example. Since these parameters are inherently non-qualitative, and extensive in their number; model explainability is understood as the process of attempting to derive meaning from learned parameters.

My argument is the following: (P1) If a scientific problem is successfully represented by schema I, then the auxiliary statements have explanatory power. (P2) If a scientific problem is successfully represented by schema II, then theory must be conjoined with prediction for facts to be explained. (C) Therefore, the critical and explanatory tendencies are necessarily interdependent. That is, if a scientific problem is represented by the critical tendency, then it is dependent on the explanatory tendency. And if a scientific problem is represented by the explanatory tendency, then it is dependent on the critical tendency. Throughout, I present why the nature and capacity of deep learning models is a representative example of C and show how it is an interesting way of interpreting both tendencies.

### Support for P1

The first premise of my argument is that when a scientific problem can be represented with the first schema; the auxiliary statements must have explanatory power for the representation to be successful.

In the same paper, (The 'Corroboration' of Theories), Putnam makes use of an example to show that a theory is in fact never used on its own to make a prediction. He states that when attempting to predict the orbit of earth; as a rudimentary example, the following statements would be assumed.

"

(I) No bodies exist except the sun and the earth.
(II) The sun and the earth exist in a hard vacuum.
(III) The sun and the earth are subject to no forces except mutually induced gravitational forces.

" (Putnam, 1979)

Interpreted with the first schema, the theory of universal gravitation conjoined with these assumptions allows for predictions to be made, it also entails that there is ambiguity, when prediction fails, as to what lead to the failure. He further suggests that the auxiliary statements will be questioned and reformulated before the theory itself would be falsified. Thus, theory does not predict alone. My point here, and with this example is that the auxiliary statements also provide explanatory support for the theory. Furthermore, I suggest that in a similar manner to the way theory alone cannot make predictions without auxiliary statements, theory alone cannot make predictions without auxiliary explanatory power. The argument is much the same as Putnam's, laws or theories are simple statements, and thus there is a necessity to conjoin them with auxiliary statements to apply them. The difference is one of emphasis, the auxiliary statements must serve to *explain* where the theory/law is applied. And as opposed to emphasizing the auxiliary statements as making predictions literally possible, the explanatory power of the auxiliary statements make the predictions *meaningful*. The essential nature of explanatory power as a cognitive value is made more evident when the predictive power of theories are equivalent.

In "Rationality and Paradigm Change" McMullin points out that contrary to the Kuhnian interpretation of the Copernican revolution, the paradigm shift from Ptolemaic theory to Copernican theory was the result of the epistemic significance of explanatory power as a cognitive value. He points out that Ptolemaic theory was only falsified after the revolution took place, and that the shift was more than a mere matter of taste (Curd & Clover, 1998). As such, both theories had a Schema I interpretation, with different auxiliary statements, but with equivalent predictive power. Additionally, the auxiliary statements were malleable enough to support predictive errors. However, the set of auxiliary statements associated with Copernican theory had greater explanatory power with regards to various related phenomena such as restricted elongation and retrograde motion. Thus, Kepler and Galileo adopted Copernican theory more eagerly.

I would like to add at this point, that I consider the epistemic value of explanatory power more significant than other epistemic values that could be present in the auxiliary statements of a schema I representation of a scientific problem. My support for this would be the claim that epistemic values can be reduced to either predictive or explanatory power. Let us take McMullins, "fertility" as an example. That is, a theory is fertile if it can generate novel research avenues. A theory that has high explanatory power and predictive power has just the same capacity to generate novel research avenues. Though I do not do a historical analysis here, it seems that the *fertility* of a theory is not an intrinsic value of a theory; but a theory becomes fertile because of other factors, such as explanatory and predictive power. A similar line of argument could be taken for Khun's *fruitfulness*. Reducing other epistemic values such as scope or simplicity can't be approached in the same manner, but I believe it is possible. *Simplicity* could be interpreted under the guise of the efficiency of the predictive and explanatory power of the theory. That is, a shorter theory would have more explanatory/predictive power per its length, as opposed to a longer more convoluted theory. A similar line of argument may be taken with regard to the cognitive value of *scope* under the guise of the reach of the theory's explanatory and predictive power.

I will now comment on the relation between my general claim and deep learning models, which becomes relevant when schematising the utilisation of these models. When the steps of gathering data, writing the code, and training a machine model are completed. Using the model could be represented by the following schema.

> Model (Parameters)
>
> <u>Input</u>
>
> Output

This schema looks remarkably similar to Putnam's critical tendency. It is worth reemphasising that the output of the model is a prediction based on the data upon which the model was trained. What is interesting is that P1 of my claim holds for this schematization of the use of machine learning models. That is, it is impossible to get an output/prediction, from the model alone. It predicts based on an initial input. For example, in large language models (LLMs) a first prompt is fed into the model (which is not necessarily seen by the user) then the LLM predicts the next $n$ tokens ($n$ depends on the context window of the model). The same happens when the user asks a question to an LLM chatbot. As such the prediction is dependent on a contextualizing/explanatory input.

<u>Support for P2</u>

I shall reconstruct Putnam's argument on why a schema II type problem *may* be dependent on a schema I type problem, and support why I think this dependence is a necessity. Putnam does defend the possible interdependence of the two schemata in his paper. More specifically, he shows with an example, how the missing auxiliary statements required to explain a given fact from a specified theory can themselves be statements of a critical tendency. The example given is that a of explaining the orbit of Uranus. Put forth as such:

" Theory: U.G.

A.S.: S1

<u>Further A.S.: ????????????</u>

*Explanandum*: The orbit of Uranus " (Putnam, 1979)

S1 represents the basic assumptions typically assumed when applying the law of universal gravitation and known planets prior to 1846. To solve the puzzle or *fill the hole* of this schema II type problem, two further auxiliary statements were needed. First (S2), the assumption that "there is one and only one planet in the solar system in addition to the planets mentioned in S1" (Putnam,1979). And second (S3), the predictive statement that given Universal Gravitation, S1 and S2; there is a planet moving along a specified orbit. S3 has the following schema I structure:

" Theory: U. G.

<u>A.S.: S1, S2</u>

Prediction: A planet exists moving in orbit O – TRUE OR FALSE?
" (Putnam, 1979)

S3 thus acts as a low-level hypothesis whose success permits the resolution of the original schema II problem. That is:

" Theory: U.G.

<u>A.S.: S1, S2, S3</u>

*Explanandum*: the orbit of Uranus " (Putnam, 1979)

This example makes it clear how a dependence of schema II representations on schema I representations is possible. Putnam mentions that these types of representations (schema II, where there is a schema I auxiliary statement) are rarely talked about by philosophers of science and that commonly, schema I representations have law-like auxiliary statements. It is debatable whether a prediction must be based on observation. It seems acceptable to admit that a low-level hypothesis that is also a law or theory has predictive power. A schema II interpretation of a scientific problem that depends on predictive auxiliary law-like statement would look like this.

> Theory
>
> AS:  ( Theory
>
>  A.S.
>
>  Prediction: A law – True or False?  )
>
> Explanandum: Fact to be explained.

Of course, one might immediately refute by saying that predictions must indeed be about observation and ask how a representation such as that above would be relevant. A response could point to the interplay of theories in the practice of science. That is, one scientist may use another's theory for the purpose of conjecture. Furthermore, a defence for the analytical connection between theoretical conjecture and statement about observation (predictions) may be explored in Carnap's Aufbau. Given these considerations, it seems reasonable for schema II type representations to use predictive law-like auxiliary statements. These A.S. would act like low-level hypotheses whose success depends on the explanatory power of the overarching schema II representation.

A parallelism can be drawn when it comes to machine learning models, more specifically with regards to extracting information from the parameters of the models. During the training process of deep learning models, no attention is generally given to the meaning of the resulting parameters. The focus is on minimalizing the loss function; that is, the difference between the predicted and the actual output after a particular training input. After the loss is considered minimized, then the model is considered trained. At this point the study of model explainability tries to derive meaning from these parameters. A successful example of such a process (for large language models) would be the clustering of vectors to general semantic categories. One among the many other methods, that does not directly manipulate the parameters, is testing various prompts to better understand how and why the model responds the way it does. As such it would schematically look something like this:

> Molel (Parameters)
>
> Input …   Expected Output – True or False
>
> Model feature to be explained

In a similar manner to schema II representations of scientific problems, successful prediction from input to output, not for model training purposes but to understand model features, support the explanatory power of the schematic representation. So far, I have shown and explained how Putnam's tendencies can be interpreted through machine learning. In the conclusion I support why I think there is significant philosophical weight to this.

Conclusion

At the core, it looks like my philosophical claim becomes one of emphasis, and otherwise is a reconstruction of many of the same movements as Putnam's in "The 'Corroboration' of Theories". If a schema I representation of a scientific problem does not have explanatory power in the auxiliary statements, it would be hard to determine what the prediction is doing exactly. The explanatory power may be minimal, such as contextualizing universal gravitation to two bodies in our solar system. But this does serve nonetheless to explain where the orbit prediction is taking place, grounding the prediction for our interpretation an understanding. Similarly, if a schema II representation of a scientific problem explains a fact derived from a theory, but the theory's auxiliary statements have no predictive power whatsoever, how do we justify the relevance of the representation? The predictive power of the auxiliary statement may also be limited or indeed abstract (as in predictive power in relation to another schema II representation), but the predictive power must be there if the representation is to be used by the scientific community.

There are a few problems with the claims I put forward. First, it may seem as though I have included a weasel word in my argument. My intent in adding the term "successfully" when first presenting the argument was to make it clear that a schematic representation of either tendency is arguably more successful because of its dependence on the other tendency. It can be put forth that it is possible to represent a scientific problem without the other tendency; however, I suggest that it is the worse for it. Second and surely more glaringly, I have suggested a rather bold equivocation; that of *theory* to *machine learning model*. I believe this is an appropriate equivocation for a few reasons.

Evidence points to continued and significant advancement of the capability of these models. The paper "Attention is All You Need", a seminal work that introduced the transformer architecture upon which major gains in LLM capabilities were achieve, was published in 2017. Today LLM Chatbots are capable enough to have become an everyday tool for many. A study published in 2024 by Porter and Machery at the University of Pittsburgh has shown that "AI-generated poetry is indistinguishable from human-written poetry and is rated more favorably". In another 2024 paper published by the Palo Alto Archetype AI research team, *Phenomenological AI Foundation Model For Physical Signals*, a model was trained on "0.59 billion samples of cross-modal sensor measurements". It was shown that without specific instruction regarding established physical laws, the model was able to predict physical phenomena on new data. Such phenomena included tracking trajectories of spring mass systems and forecasting large electrical grid dynamics.

This recent significant progress and continued investment into the infrastructure required to train machine learning models, i.e. computing facilities and power plants, point towards the continued incremental but significant progress of machine learning models. Assuming there is enough available raw data, compute, and energy; and that certain technological scalability issues are overcome; there is a reasonable amount of evidence to suggest that machine learning models will be able to provide significantly qualitatively better predictions than presently is the case. For example, a model such as that trained by the Archetype AI research team could make *better* predictions than what is possible by our current physical laws. Of course, this is only a possibility if such a discovery exists *within the data*, and that this information is not filtered out by the means by which the data is captured. The point being the limitation may be our understanding of the data, not the data itself. It is for this reason that I think that understanding and applying the parallels between Putnam's schemata and machine learning models is pertinent, and that more generally applying philosophy to machine learning is an important avenue of research. That small wrench that I hope to better understand is the idea that through an analysis of various machine learning models, model-independent objective parameters may conceivably be discovered. It would certainly be paradigm shifting if one day the talk was about the parameter-ladenness of theories.